\documentclass[prd,twocolumn]{revtex4}

\usepackage{psfrag}
\usepackage{graphicx}
\usepackage{graphics}
\usepackage{bm}
\usepackage{color}
\usepackage{verbatim,color,ulem}

\newcommand{\be}{\begin{equation}}
\newcommand{\ee}{\end{equation}}
\newcommand{\ba}{\begin{eqnarray}}
\newcommand{\ea}{\end{eqnarray}}

\newcommand{\dcom}[1]{}
\newcommand{\dnote}[1]{}

\newcommand{\gsim}{\raise.3ex\hbox{$>$\kern-.75em\lower1ex\hbox{$\sim$}}}
\newcommand{\lsim}{\raise.3ex\hbox{$<$\kern-.75em\lower1ex\hbox{$\sim$}}}

\begin{document}

\renewcommand{\thefootnote}{\fnsymbol{footnote}}


\renewcommand{\thefootnote}{\arabic{footnote}}
\setcounter{footnote}{0} \typeout{--- Main Text Start ---}

\title{The Non-universal behaviour of Cold Fermi Condensates with
Narrow Feshbach Resonances}
\author{ Chi-Yong  Lin and Da-Shin  Lee}\affiliation{
Department of Physics, National Dong Hwa University, Hua-Lien,
Taiwan 974, R.O.C.  } \author{Ray\ J.\ Rivers}
\affiliation{ Blackett Laboratory, Imperial College\\
London SW7 2BZ, U.K.}

\date{\today}
\begin{abstract}

In this paper we construct an effective field theory for a
condensate of cold Fermi atoms whose scattering is controlled by a
narrow Feshbach resonance. We show how, from first principles, it
permits a hydrodynamic description of the BEC-BCS
crossover from which the equation of state, intimately related to
the speed of sound, can be derived. Specifically, we stress the
non-universal behaviour of the equation of state at the unitary
limit of infinite scattering length that arises when either, or
both, of the range of the inter-atomic force and the scale of the
molecular field become large.
\end{abstract}


\maketitle

\section{Introduction: Scattering with Feshbach Resonances}
 Cold alkali atoms whose scattering is controlled by a Feshbach resonance can form diatomic
molecules  with tunable binding energy on applying an external
magnetic field \cite{exp_MolecularBEC}. Weak
fermionic pairing gives a BCS theory of Cooper pairs, whereas strong
fermionic pairing gives a BEC theory of diatomic molecules. The
transition is characterised by a crossover in which, most simply,
the $s$-wave scattering length $a_S$ diverges as it changes sign
\cite{exp_Crossover}.

A considerable theoretical  and experimental effort has been
expended on understanding such macroscopic quantum systems. We wish to show that, in a formalism
in which the tuning of the system by the external field is explicit because of the narrowness of the resonance, it
is straightforward to derive the semiclassical attributes of the condensates (
speed of sound, hydrodynamics, equation of state)
analytically. Our
starting point is the exemplary 'two-channel' microscopic action (in units in
which $\hbar = 1$)
 \begin{eqnarray}
S &=& \int dt\,d^3x\bigg\{\sum_{\uparrow , \downarrow}
\psi^*_{\sigma} (x)\ \left[ i \
\partial_t + \frac{\nabla^2}{2m} + \mu \right] \ \psi_{\sigma} (x)
\nonumber \\
& +& U \ \psi^*_{\uparrow} (x)\  \psi^*_{\downarrow} (x) \
\psi_{\downarrow} (x) \ \psi_{\uparrow} (x)
\nonumber \\
   &+& \phi^{*}(x) \ \left[ i  \ \partial_t + \frac{\nabla^2}{2M} + 2 \mu -
\nu \right] \ \phi(x) \nonumber \\
&-& g \left[ \phi^{*}(x) \ \psi_{\downarrow} (x) \ \psi_{\uparrow}
(x) +  \phi(x) \ \psi^{*}_{\uparrow} (x) \
\psi^{*}_{\downarrow} (x) \right]\bigg\} \label{Lin}
\end{eqnarray}
for cold fermi fields $\psi_{\sigma}$
 with spin label $\sigma = (\uparrow, \downarrow)$, which possess a {\it narrow} bound-state
 (Feshbach) resonance with tunable
binding energy $\nu$, represented by a diatomic field $\phi$ with
mass $M =2m$ \cite{theo_0,griffin,stringariRECENT,dieh}. In addition there is a simple s-wave attractive contact
interaction $U > 0$, regularised at a momentum scale $\Lambda$.
 Standard renormalisation methods \cite{rivers} allow us to remove $\Lambda $ from the formalism.
  In what follows we assume that such renormalisation has been made. The effective atomic coupling strength
  (at zero external
energy-momentum) is \be U_{eff} = U + g^2/(\nu - 2\mu) = -k_F
a_S/N_0, \label{Ueff} \ee
 comprising the contact term plus the effect of the
resonance, where  $ N_0 $ is the density of states at the
Fermi surface. Tuning $\nu - 2\mu$ to zero by the application of an external magnetic field sends $|a_S|\rightarrow \infty$. This is the so-called 'unitary' regime.

This work follows on from that of an earlier paper \cite{rivers},
 henceforth referred to as I, from which we recreate the condensate effective
 action (\ref{Leff}) in the next section.

We shall show, in greater detail than in I, that the ($T=0$)
condensate described by (\ref{Lin}) can be understood, in the
hydrodynamical approximation, as {\it two} coupled fluids, built
from the fermion-paired atomic and molecular subsystems. This is
sufficient to determine the equation of state (EOS) and speed of
sound exactly in the mean-field approximation. There are
simplifications in that a {\it single} fluid dominates in
 a) the deep BEC regime
 b) the deep BCS regime
 and c)  the unitary regime.
 When a single fluid description is appropriate the EOS can
 show several different
  allometric behaviors $p\propto\rho^{1+\gamma}$
  depending on the magnitudes of the effective range of the inter-atomic force and the length scale of the molecular field.

  In particular, when either is large,
the single fluid behaviour in the unitary regime does not show the
canonical value $\gamma = 2/3$ and we do not have
 a conformal field theory when $|a_S|\rightarrow \infty$. There has been much activity recently on
 taking advantage of  the dualities of Anti de-Sitter general relativity and conformal field theory (AdS/CFT) \cite{CFT} to use
 classical black hole physics to describe
 cold Fermi atoms in the unitary regime in those cases when the theory is conformally invariant \cite{ADS}, but our models show the limitations of  this analysis.

 Possible experimental tests are considered.

\section{Effective Actions}
  Introducing auxiliary fermion-paired  bosonic fields $ \Delta (x) = U\psi_{\downarrow}
(x)  \psi_{\uparrow} (x),\,\Delta^* (x) = U\psi^*_{\downarrow} (x)
\psi^*_{\uparrow} (x)$ renders
 ${S}$ quadratic in the fermi fields. Integrating them out \cite{rivers} enables
 us to write ${S}$ in the non-local form
  \begin{eqnarray}
 S_{NL} &=& -i\,Tr\ln {\cal G}^{-1} + \int dt~ d{\bf x}\bigg\{- \frac{1}{U}|\Delta |^2\nonumber
 \\
 &+& \phi^{*}(x) \ \left[ i  \ \partial_t + \frac{\nabla^2}{2M} + 2 \mu -
\nu \right] \ \phi(x)\bigg\},
 \end{eqnarray}
 in which
  ${\cal G}^{-1}$ is the
inverse Nambu Green function,
\begin{equation}
 {\cal G}^{-1} = \left( \begin{array}{cc}
        i \partial_t - \varepsilon         & \tilde{\Delta}(x) \\
                  \tilde{\Delta}^{*}(x) & i \partial_t +
                  \varepsilon
                  \end{array} \right)  \label{greenfun}
                  \end{equation}
 where
 $\tilde\Delta (x) = \Delta (x) - g\,\phi (x)$
  represents the two-component {\it combined } condensate (and
$\varepsilon = - \nabla^2/2m - \mu $).

In this paper we restrict ourselves to the {\it mean-field
approximation}, the general solution to $\delta S_{NL} =0$, valid
if $\phi$ is a sufficiently narrow resonance \cite{dieh,gurarie}. Our approach is a generalisation of that in \cite{aitchison},
discussed by several authors from \cite{sademelo} onwards (see Schakel \cite{schakel1}).

If we write $\Delta(x) = |\Delta(x)|\ e^{i
\theta_{\Delta}(x)}$ and $\phi(x) = -|\phi(x)| \ e^{i
\theta_{\phi}(x)}$ the combined  condensate amplitude and phase of
$\tilde{\Delta}(x) = |\tilde{\Delta}(x)| \ e^{i
\theta_{\tilde{\Delta}}(x)}$ are then determined. The action
possesses a $U(1)$ invariance under
$\theta_{\Delta}\rightarrow\theta_{\Delta} + \rm{const.}
 ,\,\theta_{\phi}\rightarrow\theta_{\phi}+\rm{const.}$,
 which
is spontaneously broken:
 $\delta\,S_{NL} = 0$ permits spacetime constant {\it gap} solutions $|\Delta
 (x)|=|\Delta_0|\neq 0$ and $|\phi (x)| = |\phi_0|\neq
 0$ (whereby $(|{\tilde\Delta} (x)|=|{\tilde\Delta}_0|\neq
 0)$ and a Goldstone boson, the (gapless) phonon.
$|{\tilde\Delta}_0|$ determines the density of states at the Fermi
surface as $ N_0 = \int d^{3} {\bf p}/ (2\pi)^3 (
|\tilde{\Delta}_{0}|^2 / 2 E_{p}^3)$, where
$E^2_{p}=\varepsilon_{p}^2 + |\tilde{\Delta}_{0}|^2$,
$\varepsilon_{p}={\bf p}^2/2m-\mu$. If $|\tilde{\Delta}_{0}| =
|\Delta_{0}| + |-g\phi_{0}|$, then $|\Delta_{0}|/|
\tilde{\Delta}_{0} | = U / U_{\rm eff} ,\,\,\,\,\, |\phi_{0}|/ |
\tilde{\Delta}_{0} | = g /[(\nu - 2\mu)\, U_{\rm eff}]$. In
addition, the system possess a gapped ('Higgs') mode.

To determine the EOS it is sufficient to consider just the
fluctuations around the gap configurations and perturb in the {\it
small} fluctuations in the scalar condensate densities
\cite{aitchison,sademelo} $\delta |\Delta|= |\Delta| - |\Delta_{0}|$
and $\delta |\phi|= |\phi|  -  |\phi_{0}|$ and their derivatives.
 We
perform a {\it Galilean invariant} long wavelength, low-frequency
expansion in space and time derivatives
 to give
\be
S_{NL}\approx S_{eff} = \int dt\,d{\bf x}\, L_{eff}
\label{SNL}
\ee
in terms of the {\it local} effective Lagrangian
density $L_{eff}$ with
elliptic equations of motion. Although $\theta_{\Delta}$ and
$\theta_{\phi}$ are not small, we assume that $\delta
|\Delta|$, $\delta |\phi|$, and $(\theta_{{\Delta}} -
\theta_{\phi})^2$ are of the same order.

 Because ${\cal G}^{-1}$ is defined in terms of ${\tilde\Delta}$ it is natural to express the local Lagrangian density
 $L_{eff}$ in terms of the phase angles $\theta_{\tilde{\Delta}}$,
$\theta_{\phi}$ and the Galilean scalar fluctuations $\delta
 |\tilde{\Delta}| = |\tilde{\Delta}| - |\tilde{\Delta}_0|$ and
$\delta|\phi|$. In fact, it is convenient to rescale $\delta
 |\tilde{\Delta}|$ to $\delta
 |\tilde{\Delta}| = \kappa\epsilon$, such that $L_{eff}$ takes the form
    \begin{eqnarray}
 L_{eff} &=& -\frac{1}{2}\rho_B
 \ G(\theta_{\phi})
 - \frac{1}{2}\Omega^2(\theta_{\tilde{\Delta}} -
 \theta_{\phi})^2\nonumber
 \\
&& -\frac{1}{2}(\rho_F^0 + 2\alpha\,\epsilon)\,
   G(\theta_{\tilde{\Delta}}, \epsilon)  + \frac{N_0}{4}\ G^2(\theta_{\tilde{\Delta}}, \epsilon)
 \nonumber\\
 && +
 \frac{1}{4}\eta X^2(\epsilon,\theta_{\tilde{\Delta}})
  -\frac{1}{4}{\bar M}^2\epsilon^2  + \frac{2g}{U}\epsilon\delta |\phi|\nonumber
  \\
  &&+ \delta |\phi| \left(\frac{\nabla^2}{4m}
+ (2\mu - \nu)\frac{U_{eff}}{ U}
 \right) \delta |\phi|,
 \label{Leff}
 \end{eqnarray}
 in terms of the Galilean
 scalar combinations
 $G(\theta ) = \dot{\theta} + (\nabla
\theta )^2/4m$, $G(\theta, \epsilon ) = \dot{\theta} + (\nabla
\theta )^2/4m + (\nabla \epsilon )^2/4m$, $X(\epsilon,\theta ) =
\dot{\epsilon}+ \nabla \theta .\nabla \epsilon/2m$ and
$\theta_{\tilde{\Delta}} -
  \theta_{\phi}$.
We have chosen $\kappa$ so that the dimensionless $\epsilon$ has the
 same coefficients as $\theta_{\tilde{\Delta}}$ in its spatial derivatives.
 On taking $g$ identically zero in (\ref{Leff}) we recover the Lagrangian of \cite{aitchison}.

  The fermion number density arising from the gap equations is $
\rho^0 = \rho_F^0 + \rho_B^0$ ,
  where
  $  \rho_F^0 = \int d^3 {\bf p} / (2\pi)^3 \ \left[ 1 -
\varepsilon_{p}/E_{p}
  \right] $
is the explicit fermion density, and $\rho_B^0 = 2|\phi_{0}|^2$ is
due to molecules (two fermions per molecule). We shall introduce
the other coefficients, which are simple momenta integrals,
 as and when they are needed.  It is the form of
(\ref{Leff}) rather than the detail that concerns us at the
moment.

\section{Equations of State}

The definition of a {\it narrow} resonance is that
\be
 \gamma_r\sim \sqrt{\Gamma/\epsilon_F}\ll 1,
\ee
where $\Gamma$ is the resonance width
and $\epsilon_F = k_F^2/2m$ is the typical atomic kinetic energy. Unless stated otherwise, we assume narrow resonances, for which the mean field approximation can be justified \cite{gurarie}.  Since the Fermi momentum $k_F$ increases as the
density $\rho$ increases
we can, in principle, make even broad resonances narrow by increasing the density. In more detail \cite{gurarie},
\be
\gamma_r = \frac{1}{(3\pi^8)^{1/3}}\frac{m^2(2\mu -\nu)^2U_{eff}^2}{\rho^{1/3}g^2}
\label{gamma}
\ee
\\
Note that $\gamma_r\propto g^2$ when $|a_S|\rightarrow \infty$, irrespective of $U$.
However, $\gamma_r$ increases as we move into the deep BEC and BCS
regimes, when $|\nu|$ is large, making the narrow resonance
approximation less valid. With this qualification, systems with
narrow resonances include $^6Li$ with $\gamma_r\approx 0.2$. On the other hand, when
$\gamma_r\gg 1$ and the narrow resonance approximation breaks down,
the model effectively becomes a one-channel model in its basic
properties.  There are then strong similarities with the
single-channel model in which $g=0$ identically, discussed by many
authors, but for which we cite \cite{aitchison} in particular.

The hydrodynamics of the system is encoded in $\theta_{\Delta}(x)$
and $\theta_{\phi}(x)$ and will have a natural realisation as two
coupled fluids. To proceed,  we ignore the density and velocity
fluctuations $\rho_{\epsilon} = -(\eta/4)
X(\epsilon,\theta_{\tilde{\Delta}})$  and ${\bf v}_{\epsilon} =
\nabla\epsilon/2m$ due to the condensate fluctuations $\epsilon$, in
comparison to $\rho^0$ and
 ${\bf\tilde v} =
\nabla\theta_{\tilde\Delta}/2m$, the condensate  velocity. The
inclusion of $\rho_{\epsilon}$ and ${\bf v}_{\epsilon}$ in the
two-fluid model would give small fluctuating short-range sources and
sinks in the fluids. All that we need for the EOS is the
hydrodynamic approximation, which coarse-grains by replacing them
with their (zero) averages.
\\
\\
 The angular Euler-Lagrange (EL) equations are then
 \begin{eqnarray}
 \frac{\partial}{\partial t}\rho_{F} + {\bf\nabla}
 (\rho_F{\bf\tilde v}) - 2{\Omega}^2 (\theta_{\tilde{\Delta}}
 -\theta_{\phi}) &=& 0,\label{hydroF}
 \nonumber
 \\
 \frac{\partial}{\partial t}\rho_{B} + {\bf\nabla}
 (\rho_B{\bf u}) + 2{\Omega}^2 (\theta_{\tilde{\Delta}}
 -\theta_{\phi}) &=& 0,\label{hydroB}
 \end{eqnarray}
 where ${\bf u} = \nabla\theta_{\phi}/2m$ and $\rho_F=\rho_F^{0} -
N_0\,G(\theta_{\tilde{\Delta}}, \epsilon) + 2\alpha\epsilon$,
 $\rho_B = 2|\phi |^2\approx \rho_B^0 +
4|\phi_{0}|\,\delta|\phi|$.
Putting these together gives
\begin{equation}
 \frac{\partial}{\partial t}(\rho_{B}+\rho_{F}) + {\bf\nabla}
 .(\rho_{B}{\bf u} +\rho_{F}{\bf\tilde v})  = 0,\label{2fluid}
 \end{equation}
 the continuity equation for two coupled fluids, as given in I.

 Eq.(\ref{2fluid}) can be written in a more transparent form.
 Whereas the explicit fermion density in molecules $\rho_B$ in (\ref{2fluid}) appears in
 conjunction with the velocity $\bf u$ of the molecular component of the
 fluid, the generalised fermion  pair density $\rho_F$ is coupled to
 the {\it combined} condensate velocity $\bf\tilde v$, rather than
 ${\bf v} = \nabla\theta_{\Delta}/2m$. However, from the definition of
${\tilde{\Delta}}$ it follows that
 $\theta_{\tilde{\Delta}} = b\theta_{{\Delta}} +
 (1-b)\theta_{\phi}$,
 where $b = |\Delta_0|/|\tilde\Delta_0| = U/U_{eff}$. In
 consequence, $ {\bf\tilde v} = b{\bf v} + (1-b){\bf u}$,
 whereby (\ref{2fluid}) can be written as
  \begin{equation}
 \frac{\partial}{\partial t}(\bar\rho_{B}+\bar\rho_{F}) + {\bf\nabla}
 .(\bar\rho_{B}{\bf u} +\bar\rho_{F}{\bf v})  = 0,\label{2fluidbar}
 \end{equation}
 where $\bar\rho_{F} = b\rho_{F}$ and $\bar\rho_{B} = \rho_{B} +
 (1-b)\rho_F$.
 That is, the effective  molecular density $\bar\rho_B$ describes point-particle bosons together with a cloud of
 fermionic Cooper pairs, which deplete the effective fermion pair density
 $\bar\rho_F$.
 Because the fluids are coupled, the condensate moves as a
 single entity with velocity $\bf\tilde v$.

The Bernoulli equations from which the EOS follows  are derived from
these EL equations on substituting for the densities. Again
neglecting $\rho_{\epsilon}$ and ${\bf v}_{\epsilon}$, the EL
equation for $\theta_{\tilde{\Delta}}$
 can be written as the simple
Bernoulli equation
\begin{equation}
 m\dot{\bf\tilde v} + \nabla \bigg[\delta h_F + \frac{1}{2}m\,{\tilde v}^2\bigg] =
 0,
 \label{Bern3}
\end{equation}
 where
 $\delta h_F = (\rho_F -\rho_F^0
- 2\alpha\epsilon)/2N_0$
  is the specific enthalpy.
 After substituting $\epsilon$ from its EL equation
 \be
 {\bar M}^2\epsilon + 2\alpha\,G({\theta}_{\tilde{\Delta}})  -  (4g/U)\delta|\phi|\approx 0
 \label{epsilon}
 \ee
 in $\delta h_F$, the enthalpy can be expressed  in terms of the density fluctuations
$\delta\rho_F = \rho_F - \rho^0_F$ and $\delta\rho_B =
4|\phi_0|\delta\phi$ as
 \be
 \delta
 h_F =  \frac{\delta p}{\rho_F^0} = K_{FF}\delta\rho_F + K_{FB}\delta\rho_B
 \label{deltadelta}
 \ee
  where
 \be
K_{FF} = \frac{{\bar M}^2}{2\big(N_0\,{\bar M}^2 + 4\alpha^2\big)},\,\ K_{FB} =
\frac{ - g\alpha}{U|\phi_0|\big(N_0\,{\bar M}^2 + 4\alpha^2\big)}.
\label{hF}
 \ee

Complementarily, the EL equation for $\theta_{\phi}$ has the form
 \begin{equation}
 m\dot{\bf u} + \nabla \bigg[\delta h_B + \frac{1}{2}m\,{u}^2 -\frac{1}{16m\rho_B^0}\nabla^2\delta\rho_B\bigg] =
 0,
 \label{Bern6}
\end{equation}
where $h_B$ permits the decomposition
 \be
 \delta h_B =  \frac{\delta p}{\rho_B^0} = K_{BF}\delta\rho_F + K_{BB}\delta\rho_B.
 \label{deltadelta2}
 \ee
 As required, $K_{BF} = K_{FB}$ and
\be
 K_{BB} = \frac{(\nu -
2\mu)}{8|\phi_0|^2}\frac{U_{eff}}{U} - \frac{g^2}{2U^2|\phi_0|^2}
\frac{N_0}{4\alpha^2 + N_0 {\bar M}^2}
 .\label{KBB}
 \ee

 The $\nabla^2\delta\rho_B \propto \nabla^2\delta|\phi|$ term in (\ref{Bern6}) is just as we
would expect from a theory of a pure bosonic gas. In the
hydrodynamic approximation such derivatives of $\delta|\phi|$ are
also ignored comparatively, and the resulting equation
 \begin{equation}
 m\dot{\bf u} + \nabla \bigg[\delta h_B + \frac{1}{2}m\,{u}^2\bigg] =
 0,
 \label{Bern7}
\end{equation}
is taken in conjunction with (\ref{Bern3}) in determining the EOS.

It follows from (\ref{deltadelta}) and (\ref{deltadelta2}) that
  \be
 \delta\rho_F(\rho_F^0 K_{FF} - \rho_B^0 K_{BF}) =
 \delta\rho_B(\rho_B^0 K_{BB} - \rho_F^0 K_{FB}).
  \ee
 We have learned from (\ref{2fluid}) and (\ref{2fluidbar}) that, however it may be partitioned,
 the total fermion density is $\rho_F + \rho_B$, whose fluctuation is $\delta\rho = \delta\rho_F + \delta\rho_B$.
 Eqs. (\ref{deltadelta}) and (\ref{deltadelta2})
 then collapse to give the EOS of the condensate as
 \be
 \frac{dp}{d\rho} = \frac{\rho_F^0\rho_B^0 (K_{FF}K_{BB} - K_{FB}K_{BF})}{(\rho_F^0 K_{FF} +\rho_B^0 K_{BB}
 - \rho_B^0 K_{BF} - \rho_F^0 K_{FB})},
  \label{EOS}
 \ee
 where we remember that $K_{FB} = K_{BF} < 0$. Eq.(\ref{EOS}) is the key equation of this paper.

As a very good check on our calculations, the time derivatives of
the EL equations
 for $\theta_{\phi}$ and $\theta_{\phi}$ determine the dispersion relations
 of the modes. It follows directly that the speed of sound
 $v$
 is given as
 \be
 v^2 = \frac{\rho^0}{m}\bigg(\frac{K_{FF}K_{BB} - K^2_{BF}}{K_{FF} + K_{BB} - 2K_{BF}}\bigg).
 \label{v2}
 \ee
Neither $dp/d\rho$ nor $v^2$ depend on the coefficients $\eta$,
$\Omega^2$ in (\ref{Leff}) or the scaling parameter $\kappa$ and we
use the results of \cite{rivers} to identify the coefficients in the
$K$s as $\alpha = \int d^3 {\bf p} / (2\pi)^3
(|\tilde{\Delta}_0|\varepsilon_{p}/ 2E_{p}^3)$ and ${\bar M}^2 =
2(2/U - \beta) > 0$, where
 \be
  \beta = \int \frac{d^3 {\bf p}}{
(2\pi)^3} \bigg[\frac{\varepsilon_{p}^2}{ E_{p}^3} -
\frac{1}{ ({\bf p}^2/2m)}\bigg] < 0.
 \ee
  We assume that all
the parameters in $dp/d\rho$ and $v^2$ have been renormalised as
in I.
On substituting this and the $K$s above into $v^2$ of (\ref{v2})
we do, indeed, recover the results of I.
Unfortunately, although $dp/d\rho$ is given in terms of
 straightforward momenta integrals, in general it bears no simple relationship to $v^2$. This is not surprising since the system comprises two
 coupled fluids.


We conclude this section with a further comment on the role of the
order parameter  fluctuations $\epsilon$ and $\delta|\phi|$
(deviations from homogeneous gap solutions). The coarse-graining
described above is not only sufficient to obtain the EOS, but
obligatory within the approximation (\ref{SNL}), that corresponds to
taking only the lowest relevant field derivatives, in that the
fluctuations are too short-range or too fast \cite{aitchison} to be
properly accounted for by it.

To go beyond the EOS requires an understanding of the fluctuations
of the gapped 'Higgs' mode mentioned earlier, built from order
parameter fluctuations. (This is trivially so in the limiting case
of $g=0$ for which the Higgs field is just $\epsilon$
\cite{aitchison}.)  Equally, the dynamics of the Higgs mode cannot
be obtained reliably from the approximation (\ref{SNL}) and we need
to incorporate higher-order field derivatives non-perturbatively  to
describe it (e.g. see \cite{barankov}).


For our purposes we can ignore the Higgs mode and we revert to our
results of (\ref{EOS}) and (\ref{v2}).

\section{Single fluid regimes}

 With the qualifications above, there are circumstances in which we have a description in terms of a {\it single} fluid. Most simply, in the
deep BCS regime (where $\rho_B\approx 0$) and the deep BEC regime
(where $\rho_F\approx 0$), the system of (\ref{2fluid}) behaves as a
 {\it single} fluid, of Cooper pairs or molecules respectively.
A characteristic of these extremes is that the terms that are negligible in $dp/d\rho$ are also negligible in $v^2$,
 whereby
 we get the simple result $dp/d\rho \propto m v^2$, as we shall now show.

 \subsection{The BCS regime}

 \noindent In the deep BCS regime $\rho_B^0 = 2|\phi_0|^2$ vanishes, as does $\alpha$ because of particle-hole symmetry.
 As a result, $K_{BB}$ becomes very large
 and
  \be
 \frac{dp}{d\rho} \approx \rho^0_F K_{FF}\approx\frac{\rho}{2N_0}\approx m\,v^2.
 \ee
 The EOS follows directly. At the Fermi surface ($\mu\approx \epsilon_F$) the density of states $N_0\propto m\rho^{1/3}$
  whence
  \be
 \frac{dp}{d\rho} \propto \mu \propto \rho^{2/3}\label{EOS1}
\ee
 or, equivalently, $p\propto \rho^{1 +\gamma}$
 where $\gamma = 2/3$. There is no pressure from molecules in this limit. This is in agreement with
 the work of other authors (e.g. \cite{stringari2002,heiselberg})

\subsection{The BEC regime}

 The deep BEC regime (large negative $\nu$) is characterised by
 small $N_0$, large $\alpha$, and $M^2\approx 4/U$. Unlike the BCS
 regime, $K_{FF}$ dominates the Ks, so that
 $v^2\approx \rho_{B} K_{BB}/m$.
 Inspection shows that
 \be
  \frac{dp}{d\rho} \approx \frac{1}{2}m\,v^2,
  \label{BECU}
 \ee
 since
 $\rho_F^0 K_{FF}\approx \rho_B^0 |K_{BF}|$.
 We have shown in I
  that in the BEC regime $v^2\propto \rho$, giving us the EOS
 \be
 p\propto \rho^{1 +\gamma}
\hspace{0.5cm}\mbox{with}\hspace{0.5cm} \gamma = 1.
 \label{allo}
 \ee
 This is the expected value of $\gamma$, corresponding to a
 molecular condensate with repulsive interaction \cite{astrak}.

\subsection{The Unitary limit}

The extremes above permit relatively trivial single-fluid descriptions.
 More important is
 the  'unitarity' limit
describing the central region $\nu=2\mu$ where
$|a_S|\rightarrow\infty$. This is described by a single fluid, not
because $\rho_B$ or $\rho_F$ vanishes, but because $b
=U/U_{eff}\rightarrow 0$. This is essentially a one-channel system
\cite{ho},  represented by the {\it single} fluid ${\bar\rho}_F =
0$, ${\bar\rho}_B = \rho$ in terms of the
 $\bar\rho\,$s of (\ref{2fluidbar}).  We stress that the divergence of $|a_S|$ is not a signal of singular behaviour. The $K$'s that define the behaviour of the system vary continuously as we pass from the BCS to BEC regimes through the unitarity regime, and vice-versa (e.g. see (\ref{KBB})), for which perturbative methods such as ours are valid for narrow enough resonances (e.g. see Fig. 4 of \cite{gurarie}).

 In addition to the scattering length $a_S$, the general model possesses two other important length scales, the effective range of the force $r_0$ where $\gamma_r\sim\hbar/k_F|r_0|$ as well as the length scale $ \xi= ( | 4m \, (2\mu-\nu) \, U_{eff}/U |  )^{-1/2}$ for the molecular field.   In general  both $ r_0 = O(g^{-2})$ and $\xi = O((g^2/U)^{-1/2})$ are small near the unitary limit for broad resonances with large resonance coupling $g$, and
 the only effective dimensionless parameter is $k_F|a_S|$.  On it diverging, the theory shows universal behaviour, with conformal invariance. In principle, AdS/CFT duality then enables us to convert the difficult strong coupling calculations needed to describe the system into more tractable weak coupling boundary calculations for classical black holes \cite{CFT}. However, for  narrow resonances $|r_0|$ becomes large and  this universal behaviour is broken. Further, for small $g^2/U$ we see that $\xi$  becomes large, again leading to the breakdown of universal behaviour.

 Nonetheless, in many regards there is still {\it de facto} universality for realistic systems [see Figs. 30 of Chen {\it et al.} \cite{chen}) for many observables. To see deviations from non-universality the best way is to look for the deviation of the EOS from its canonical behaviour.

 After UV renormalisation, the gap equation at this
limit becomes
 \begin{equation}
 0 = \frac{1}{U_{eff}} =  \int  \frac{d^3 \bf p}{(2\pi)^3}
\bigg[\frac{1}{2E_{p}}-\frac{1}{ 2( {\bf p}^2/2m)}\bigg].
\label{aS}
\end{equation}
 This fixes
 \be
 |\tilde\Delta_0| = c \mu \, ,
 \label{unitcond}
 \ee
 where $c\approx
1.16$. In turn, this relates the density $\rho$ to $\mu$ as
\vspace{-0.2cm}
 \be
 \rho = \rho_B + \rho_F = \frac{2 c^2}{g^2}\mu^2 +
 \frac{c_1}{2\pi^2}(2m\mu )^{3/2}
 \label{unitrho}
 \ee
 where the first and second terms are $\rho_B$ and $\rho_F$
 respectively and $c_1\approx 1.47$.

 Although we have constructed the model with narrow resonances in mind the formalism permits extension to broad resonances \cite{aitchison} and we consider all possibilities.
  In the first instance, for broad resonances ($\gamma_r \gg 1$), for which $k_F|r_0|\ll 1$, $\rho \approx \rho_F \gg \rho_B$, whereas for
narrow resonances ($\gamma_r \ll 1$), for which $k_F|r_0|\gg 1$, $\rho \approx \rho_B \gg \rho_F$.

In determining the EOS we stress that, in the unitary limit, the
resonance width is {\it independent} of $U$.
\subsubsection{Broad resonances 
}
The canonical conformal symmetry in the unitarity regime arises
when both $k_F|r_0|\ll 1$ and $k_F\xi\ll 1$ or, equivalently, we
have {\it large} resonance coupling $g$,
 with $ g^2 \gg  U\mu \, ,\mu^2/( 2m \mu)^{3/2} $.
 For such values $\rho_F K_{FF}\ll \rho_B K_{BB}, \rho | K_{BF}|$, and we recover the 'universal'
behavior
 \be
 \frac{dp}{d\rho}\propto \mu \propto \rho^{2/3},
 \ee
 or $p\propto \rho^{1 +\gamma}$ with $\gamma = 2/3$, as in
the BCS regime.

Although our approximation
is not wholly reliable in that, for  broad resonances, the renormalization of the
molecular boson is expected to contribute sizable corrections to
the EOS in such a strongly coupled regime~\cite{astrak}, this is the correct value \cite{stringari2002},
observed experimentally (see Fig. 15 of \cite{stringari2002}).

For broad resonances, $\mu \sim k_{F}^2 / 2m$, whereby $k_F\xi  \sim (U\mu/g^2)^{1/2}$.
For relatively {\it smaller} $g$ ($ U \mu \gg g^2 \gg \mu^2/(
2m \mu)^{3/2}$), even though $|r_0|$ remains small, conformal invariance is broken by $k_F\xi\gg 1$. In these circumstances $\rho_F K_{FF}\gg \rho_B K_{BB}, \rho | K_{BF}|$,
whereby
 \be
  \frac{dp}{d\rho}\propto \mu^{0} \propto \rho^{0}.
 \ee
giving the EOS  $p\propto \rho^{1 +\gamma}$ with $\gamma =0$.

\subsubsection{Narrow resonances 
}

When the resonance is narrow 
  the mean field approximation is robust \cite{gurarie}.  There is no contradiction in
 having
 {\it strong} self-interaction
$U$  ($U \gg \mu/(2 m\mu)^{3/2} \gg g^2/\mu$), when the conformal
symmetry, already broken by $k_F|r_0|\gg 1$, is further broken with
\be
 k_F\xi = \bigg(\frac{g^2}{U\mu}\bigg)^{-1/2}\bigg(\frac{\mu}{(2m\mu)}\bigg)^{1/3}\bigg(\frac{g^2}{\mu}\bigg)^{-1/3}\gg 1.
\ee
In this case we find $\rho_F K_{FF}\gg \rho |
K_{BF}| \gg \rho_B K_{BB}$, giving \vspace{-0.2cm}
 \be
 \frac{dp}{d\rho}\propto \mu^{0} \propto \rho^{0}.
 \ee
 That is, $p\propto \rho^{1 +\gamma}$ where $\gamma = 0$. This is
 the behaviour of a free Bose gas, expected in this limit
 \cite{nishida}.
 \\
 \\
 However, for relatively {\it weak} self-interaction ($ U \, , g^2/\mu \ll \mu/(2 m\mu)^{3/2} $), with smaller
 $k_F\xi$ than previously, we find
 $\rho_B K_{BB}\ll \rho_F K_{FF},
\rho |K_{BF}| $,  leading to
\be
 \frac{dp}{d\rho}\propto\mu^{1/2} \propto \rho^{1/4}.
 \ee
 That is, $p\propto \rho^{1 +\gamma}$ where $\gamma = 1/4$.
\\
\\
As we anticipated earlier, in all of the cases listed above it
happens that $dp/d\rho \propto m v^2$, and we can equally well
read off the behaviour from that of
 $v^2$.
Away from these extremes we see, from (\ref{unitrho}), that an
allometric representation is not justified, although the effective
exponent $\gamma_{eff} = d(\ln dp/d\rho)/d \ln\rho$ will interpolate
between these values.

 \section{Experimental tests}

 The best experimental test of our non-canonical equations of state is to determine the EOS exponent $\gamma$ directly by observing the expansion of the condensate in elongated traps on removing the potential \cite{giorgini,stringari2002}. This is governed by the hydrodynamical equations and to see how this carries over to our model we need to examine the single fluid nature of the unitary regime in more detail.
\\
\\
 Consider an  elongated axially symmetric harmonic trap with frequencies $\omega_x = \omega_y = \omega_{\bot}$ and $\omega_z$ ($\omega_{\bot} > \omega_z$).
 When there is simple allometric behaviour $\delta h\propto \rho^{\gamma}$, the density can take the scaling form
 \be
  \rho(x,y,z,t) = \frac{1}{\prod_j b_j}\rho_0\bigg(\frac{x}{b_x},\frac{y}{b_y},\frac{z}{b_z},t\bigg),
 \ee
 where $b_x = b_y = b_{\bot}$.
The continuity and Bernoulli equations then reduce to \cite{stringari2002}
 \be
 {\ddot b}_i + \frac{\omega_i^2}{b_i\prod_jb_j^{\gamma}} = 0.
 \ee

 The {\it aspect ratio}, the relative transverse to longitudinal expansion of the condensate,
 \be
  \frac{R_{\bot}(t)}{R_z(t)} =  \frac{R_{\bot}(0)}{R_z(0)}\frac{b_{\bot}(t)}{b_z(t)}
 \ee
 can be measured where $R_{\bot}$ and $R_z$ are radial and axial radii respectively.
 \begin{figure}
\centering
\includegraphics[width=\columnwidth]{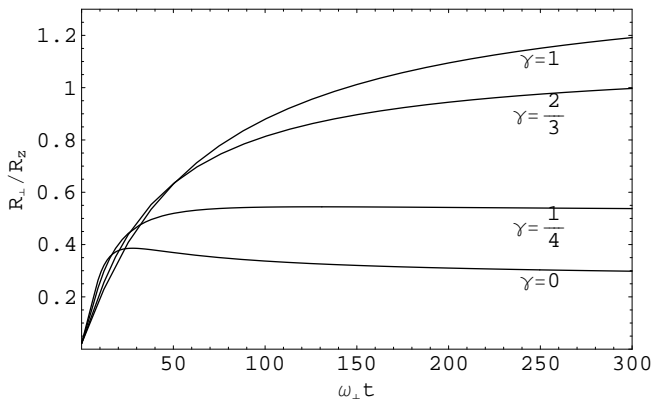}
 \caption{The aspect ratio as a function of the time for the expansion of the condensates with different allometric behaviours for its EOS when $\lambda= \omega_{\bot}/\omega_z=0.1$. The 'universal' behaviour  $\gamma = 2/3$ corresponds to strongly coupled broad resonances. For narrow resonances in the unitary limit $\gamma$ interpolates between $\gamma = 0$ and $\gamma = 1/4$   and $\gamma = 1$ arises in the BEC regime.} \label{fig1}
\end{figure}
In Fig.~(\ref{fig1}) we show this ratio for the values of $\gamma$
 derived above (in the unitarity and extreme BEC and BCS regimes) for $\lambda = \omega_{\bot}/\omega_z = 0.1$.
 It is evident that the hydrodynamical effect is greater in the direction of larger density gradients, as anticipated.
For a larger value of the exponent $\gamma$ on EOS, it is found that
the shape of the condensates at asymptotical times changes more
dramatically. The non-canonical behaviour interpolating between
$\gamma = 1/4$ and $\gamma = 0$ for narrow resonances should be the
easiest to identify.
 \\
 \\
\section{Conclusions}

We have seen that an explicitly tunable two-channel model for cold
Fermi gases with a {\it narrow} Feshbach resonance has a
two-component order parameter. With each complex component having
its own phase it is not surprising that, for the purposes of its
EOS, the system can be described in a very transparent way by two
coupled fluids.

The system can be driven from one of Cooper pairs to one of tightly
bound diatomic molecules by applying an external magnetic field. The
single fluid limits of the deep BEC and BCS regimes, in which either
the densities of Cooper pairs or molecules are zero, are familiar.
Our emphasis here has been on the very different single-fluid limit
of the unitary regime, at which the scattering length diverges, in
which neither density vanishes. For this we have shown that the EOS,
with an exponent interpolating between $\gamma = 0$ and $\gamma =
1/4$, will differ strongly from the canonical behaviour $\gamma =
2/3$ of a conformal field theory, according as one, or both, of the
effective range of the inter-atomic force and the length scale of
the molecular field become large. This can be confirmed by
measurements of the aspect ratio for elongated condensates.

We conclude with an observation on formalism. Although the two-fluid
description is very natural, the hydrodynamic equations can be
reformulated as Gross-Pitaevskii (GP) equations for coupled complex
GP fields   $\Psi_F = \sqrt{\rho_F}
e^{i\theta_{\tilde{\Delta}}}/\sqrt{2}$ and $\Psi_B = \phi
=\sqrt{\rho_B} e^{i\theta_{\phi}}/\sqrt{2}$, one field for each
density and phase. However, since phases are field logarithms, this
coupling is unmanageably logarithmic in the fields, in general. It
is only when a single-fluid description is possible that the GP
formalism is useful e.g. in the BEC regime, when
$\Psi_F\propto\tilde{\Delta}$ \cite{rivers}.

A further case for which the GP formalism is useful, that we shall
consider elsewhere \cite{rivers2}, is for the idealised situation in
which $U=0$, discussed in detail in \cite{gurarie}. In this case we
have a {\it single}-fluid description through the whole BEC-BCS
regime or, equivalently, a single GP equation. The EOS has $\gamma =
1/4$.
 \\
\\
\noindent RR would like to thank the Academia Sinica, Taipei and
National Dong Hwa University, Hua-Lien, for support and
hospitality, where much of this work was performed. The work of
DSL and CYL was supported in part by the National Science Council
and the National Center for Theoretical Sciences, Taiwan.

\end{document}